\documentclass[12pt]{iopart}
\usepackage{graphicx, lipsum}
\usepackage{float}
\usepackage{wrapfig}
\restylefloat{figure}
\begin{document}

\title{Physionary - a scientific version of Pictionary}

\author{N. Poljak, A. Bosilj, S. Brzaj,  J. Dragovic, T. Dubcek, F. Erhardt, M. Jercic}
\address{Physics department, University of Zagreb}

\begin{abstract}
We describe a variant of the popular game ``Pictionary'' based on the terms used in elementary and high school physics. We believe that the drawing of physical terms helps students develop a deeper understanding of physical concepts behind them, as well as apply them in everyday situations.
\end{abstract}

\mbox{}
\hskip -9mm
This is the version of the article before peer review or editing, as submitted by an author to Physics Education. IOP Publishing Ltd is not responsible for any errors or omissions in this version of the manuscript or any version derived from it. The Version of Record is available online at 10.1088/1361-6552/aadde4.

\section{Introduction}

The Croatian Physical Society organizes the annual Summer school for young physicists [1], intended to reward elementary and high school students for their accomplishments on the national physics competitions. The schools typically consists of half-day lectures combined with workshops, experiments or games that take place during leisure hours. For the last three editions of the summer schools, we developed a variant of the popular game "Pictionary" as a small competition for the students. ``Physionary'', as we named the variant, has proven to be very successful in entertaining the students, not only during the evenings intended for the competition, but also during the rest of leisure time.
\section{Game description}
``Physionary'' is a game loosely derived and expanded from the commercially-distributed ``Pictionary''. A similar game was developed earlier for University biology students [2], however, we expanded further on the game since we found it beneficial to do so. 

The students are divided into groups based on their age and are given a number of cards, each containing 6 terms from elementary or high school physics. The terms are taken from indices of physics curricula or physics manuals and divided into 5 sets of cards, one for each high school grade and one for elementary school grades. A dice is thrown to randomly select the ordinal number of the term on the card. A one minute timer is started and one of the students from each group is required to draw the term found on his/her card, while the rest of the group has to guess what term it is. If the team accomplishes this within the given time allotment, they receive a point. Several rounds are played this way before the pace of the game is then made faster by decreasing the available time for drawing. After a predetermined number of rounds played in this manner, the second phase of the game begins. In this phase, the students don't draw the terms but instead try to ``act out'' the term given on their cards, as it is usually done in charades. This is also done within a given time allotment, typically set to a minute.

This game turned out not to be only entertaining, but also highly educational. Sometimes, a specific term may not be recognized by some of the members of a certain group. However, we noticed that the said term is quickly taken in by those members, as evidenced by their future recognition of that, as well as similar terms. Once the students start to communicate concepts pictorially, they move from their definitions and try to use everyday examples to convey them to their group. We have also noticed that a sense of connection between various terms is formed, since the students find it beneficial to explain a new term with the help of terms that they have already drawn on paper - they simply circle the term that was already guessed by the group. Finally, the students were often found drawing graphs and diagrams, which is a skill they need to develop in physics, but are often not motivated to do so.
\begin{wrapfigure}[16]{r}{0.3\textwidth}
\begin{center}
\includegraphics[width=0.28\textwidth]{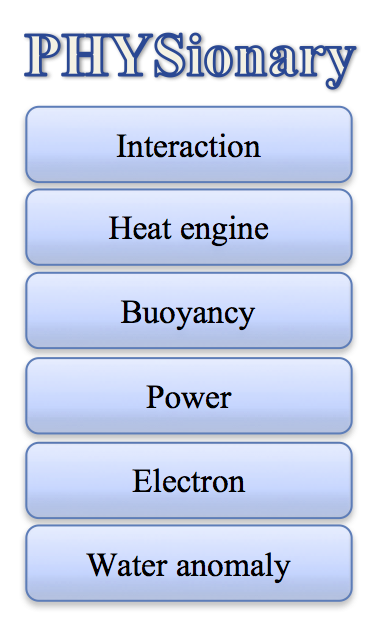}
\end{center}
\caption{An example of a playing card.}
\end{wrapfigure} 

\section{Conclusion}
We have expanded on the known popular party games to create an effective and entertaining physics learning tool. The skills developed during game play seem to be beneficial to the students and the terms they are required to draw or act out are taken from their curricula. The students seem to develop another important skill during the game - using simple physical concepts in everyday situations, which is a skill they are most often found lacking. In what follows, we present a selection of 10 terms (out of 850) from each of the 5 sets of cards. We have limited the selection to 10 terms since we expect that different countries will have differing curricula, so we thought it best that everyone interested made their own sets of cards.

\section*{References}
[1] http://www.hfd.hr/ljetna\_skola/\newline
[2] Kathleen A. Parson, John S. Miles, ``Bio-Pictionary -- a scientific party game which helps to develop pictorial communication skills'',  Journal of Biological Education, 28:1, 17-18, DOI:10.1080/00219266.1994.9655358

\newpage
\begin{center}
\begin {table}[H]
\begin{tabular}{ |c|c|c|c|c| } 
 \hline
 \bf elementary & \bf 1st grade &\bf 2nd grade &\bf 3rd grade &\bf 4th grade \\
 \hline 
 heat insulator & joule & diffusion & diffraction & plasma \\
 surface area & frequency & capacitor & rainbow & atom \\
 power & dynamics & inductivity & standing wave & antiparticle \\
 mass & buoyancy & linear expansion & intensity & semiconductor \\
 sliding friction & unit & Lorentz force & sound & fractal \\
 Solar energy & fluid & work & rotation & boson \\
 pulley & projectile motion & ideal gas & length contraction & red giant \\
 molecule & cosmic speed & insulator & phase & mass defect \\
 cavity & energy & interaction & lens & butterfly effect \\
 electricity & Galileo & isobar & light guide & quark \\
\hline
\end{tabular}
\caption {A sample from the terms given on the ``Physionary'' cards, sorted in 5 classes according to students' grade.} \label{tab:title} 
\end {table}
\end{center}
\end{document}